\documentclass[aps,prd,amssymb,superscriptaddress,footinbib,showpacs,twocolumn]{revtex4} 
\usepackage{amsmath}
\usepackage{amsfonts}
\usepackage{graphicx}
\usepackage{epsfig}
\usepackage{subfigure}
\usepackage{bbm}
\usepackage{psfrag}
\usepackage{pstool}

\newcommand{\ba}{\begin{align}}
\newcommand{\ea}{\end{align}}
\newcommand{\be}{\begin{equation}}
\newcommand{\ee}{\end{equation}}
\newcommand{\bea}{\begin{eqnarray}}
\newcommand{\eea}{\end{eqnarray}}
\newcommand{\ls}{\left[}
\newcommand{\rs}{\right]}
\newcommand{\lr}{\left(}
\newcommand{\rr}{\right)}

\newcommand{\pa}{\partial}
\DeclareMathOperator\arctanh{arctanh}
\DeclareMathOperator\arcsinh{arcsinh}

\begin{document}
\title{Twin paradox with macroscopic clocks in superconducting circuits}
\author{Joel Lindkvist}
\affiliation{Microtechnology and Nanoscience, MC2, Chalmers University of Technology, S-41296 G\"oteborg, Sweden}
\author{Carlos Sab{\'\i}n}
\affiliation{School of Mathematical Sciences,
University of Nottingham,
University Park,
Nottingham NG7 2RD,
United Kingdom}
\author{Ivette Fuentes}
\affiliation{School of Mathematical Sciences,
University of Nottingham,
University Park,
Nottingham NG7 2RD,
United Kingdom}
\author{Andrzej Dragan}  
\affiliation{Institute of Theoretical Physics, University of Warsaw, Hoza 69, 00-049 Warsaw, Poland}
\author{Ida-Maria Svensson} 
\affiliation{Microtechnology and Nanoscience, MC2, Chalmers University of Technology, S-41296 G\"oteborg, Sweden}
\author{Per Delsing} 
\affiliation{Microtechnology and Nanoscience, MC2, Chalmers University of Technology, S-41296 G\"oteborg, Sweden}
\author{G\"{o}ran Johansson}
\affiliation{Microtechnology and Nanoscience, MC2, Chalmers University of Technology, S-41296 G\"oteborg, Sweden}

\begin{abstract}
We propose an implementation of a twin paradox scenario in superconducting circuits, with velocities as large as a few percent of the speed of light. Ultrafast modulation of the boundary conditions for the electromagnetic field in a microwave cavity simulates a clock moving at relativistic speeds. Since our cavity has a finite length, the setup allows us to investigate the role of clock size as well as interesting quantum effects on time dilation. In particular, our theoretical results show that the time dilation increases for larger cavity lengths and is shifted due to quantum particle creation.
\end{abstract}
\maketitle
\section{Introduction}
Einstein's theory of relativity \cite{specrel,relativityrindler} leads to the twin paradox, in which a twin traveling at high speeds in a spaceship ages more slowly than her sibling, who stays at rest. Although constant motion is relative, the paradox is resolved by considering the acceleration experienced by the moving twin, breaking the symmetry. The fact that moving clocks tick slower is called time dilation, and it has been tested experimentally to high accuracy by observing decay rates of particles moving at relativistic speeds through the atmosphere \cite{rossi}, or in an accelerator storage ring \cite{bailey}. Another approach for verification is based on state-of-the-art clocks, where more modest speeds are enough to create measurable time differences. Such experiments include sending atomic clocks with commercial jets on east- and west-bound paths around the world \cite{hafele} and, very recently, in a ground-based laboratory where the speed of the moving-ion clock was only 10 m/s \cite{chou}.

Cutting-edge experiments in Circuit Quantum Electrodynamics (cQED) \cite{blais,reviewdevoret}, where quantum optical effects are investigated in the interaction of artificial atoms with one-dimensional electromagnetic fields, have now reached new experimental regimes beyond standard matter-radiation interactions. In particular, it has recently been suggested that it should be possible to observe relativistic quantum effects \cite{reviewnori, reviewjohansson,johanssonjohansson}, by ultrafast modulation of the boundary conditions experienced by the electromagnetic field. This enabled the experimental observation of the dynamical Casimir effect \cite{wilson} -a long-sought theoretical prediction of Quantum Field Theory- opening a new avenue to explore relativistic effects in quantum technologies \cite{friis}. Experiments in the overlap of quantum theory and relativity are of great relevance since we lack understanding about how the theories can be unified.  

In this paper, we propose a lab based experiment in which the twin paradox can be simulated with velocities approaching 2 \% of the speed of light. By ultrafast modulation of the electric length of a superconducting cavity, the electromagnetic field inside the cavity experiences similar boundary conditions as in a cavity moving at relativistic speeds \cite{friis}. Initiating the field inside the cavity in a coherent state, the phase of this state can be used as the pointer of a clock. We show that for state-of-the-art experimental parameters, the phase shift between the twin cavities can be as large as 130 degrees, which is clearly in the measurable regime. 

Unlike previous setups, our scheme addresses the effects of time dilation in relativistic quantum fields. While previous studies assumed the clock to be pointlike, in our approach the clock has a length of more than 1 cm, leading to a measurably different time dilation. In that sense, this is the first proposal to test the twin paradox with macroscopic quantum systems.  This is interesting since a pointlike clock is only affected by the instantaneous velocity and therefore can only be affected indirectly by acceleration. However, acceleration directly affects a quantum field contained in a cavity. The acceleration of the cavity's boundary conditions gives rise to the dynamical Casimir effect \cite{moore,wilson}, a genuine quantum effect where motion induced particle creation and mode-mixing among the field modes are predicted to be observable \cite{bsgates}. This enables us to address further questions in the overlap of quantum theory and relativity such as the study of new quantum effects on finite size relativistic clocks. Indeed our theoretical analysis shows that the dynamical Casimir effect and the spatial extension affect the rate of the clock, i. e. time dilation. We find that time dilation increases with the length of the superconducting cavity. In other words, the travelling twin ages less if his clock is larger. Particle creation gives rise to a small shift in the time dilation, highly dependent on the details of the trajectory. These effects show up as corrections to the standard time dilation seen by a pointlike clock.

Using the setup we propose, the time dilation effects predicted in the twin paradox, as well as the effects of clock size, can be readily demonstrated in accessible parameter regimes. Currently, however, we will not be able to reach the regimes (involving velocities as large as 25 \% of the speed of light) required to demonstrate the effects of the dynamical Casimir effect on time dilation predicted in this paper.  While these regimes have already been achieved in an experiment using a single mirror in harmonic motion \cite{wilson}, it is more challenging to mimic a cavity of constant proper length moving in those regimes. However, given the accelerated rate at which experimental advances in cQED have developed, we expect that in the near future it will be possible to confirm our predictions concerning particle creation as well.

\section{The twin paradox with cavities}
\begin{figure}[t]
\centering
\includegraphics[width=0.9\columnwidth]{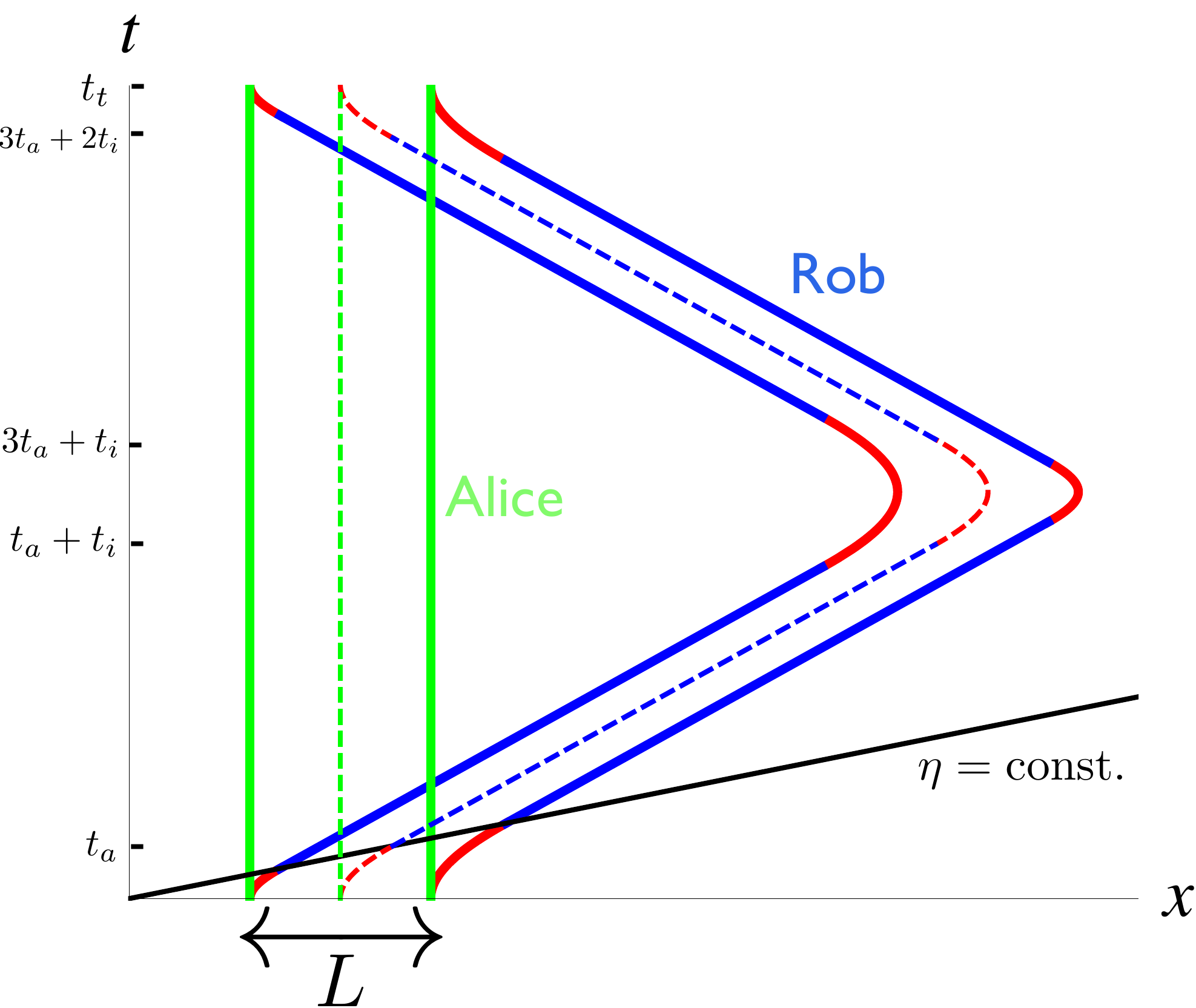}
\caption{\textbf{Cavity trajectories.} Minkowski diagram showing the cavity trajectories in the lab frame. Alice's cavity (green) stays static. The mirrors of Rob's
cavity move along trajectories composed of segments of constant proper acceleration (red) and inertial motion (blue). The dashed trajectories are those of Alice and Rob themselves and the black line is a line of constant Rindler time. Both cavities have the same proper length $L$.} 
\label{fig:Trajectory}
\end{figure}

To describe the twin paradox scenario we consider two different observers, Alice and Rob, in $1+1$-dimensional Minkowski spacetime. Alice will be inertial and stay static with respect to our lab frame, with Minkowski coordinates $(t,x)$. Rob, on the other hand, will undergo a round trip starting and ending at rest with respect to Alice, at the same spacetime point. We study a simple example of such a trip, composed of four accelerated segments and two segments of inertial motion (see figure \ref{fig:Trajectory}). During each accelerated segment, Rob moves with constant proper acceleration $a$. In the lab frame this corresponds to movement along a hyperbola in the $(t,x)$-plane. We let the duration of each segment, in the lab coordinates, be equal to $t_a$. During the inertial segments, Rob moves with a constant velocity that is set by $a$ and $t_a$ and we denote the duration of these segments by $t_i$. Thus, Rob's trajectory is completely described by $a$, $t_a$ and $t_i$. In the lab frame, the duration of the trip is $t_t\equiv 4t_a+2t_i$.

In order to compare their elapsed proper times, Alice and Rob need to carry some form of clocks. For this, we will use cavities containing quantized one-dimensional electromagnetic fields. The cavities are of constant proper length, i. e. length measured by a comoving observer. The idea is to prepare the cavities in identical coherent states. After the trip, the phase shifts in the two cavities are determined and these are used as a measure of the elapsed proper times.

In its rest frame, the cavity is constructed by inserting two perfect mirrors separated by a distance $L$. We imagine Alice and Rob sitting at the center of their respective cavities, each of proper length $L$.  When Rob moves with constant velocity, his cavity is shorter in the lab frame due to length contraction. Thus, during the accelerated segments, the two mirrors must move with different proper accelerations in order for the proper length of the cavity to stay constant. More precisely, they need to move along different hyperbolas in the $(t,x)$-plane. One of the mirrors moves with greater acceleration than Rob but for a shorter time, and vice versa for the other mirror (see figure \ref{fig:Trajectory}).
\newline
\\
For an inertial observer, a 1D electromagnetic field $\phi$ defined on a Minkowski spacetime background obeys the wave equation
\be
(\pa_t^2-c^2\pa_x^2)\phi=0,
\label{KGeqn}
\ee
where $c$ is the speed of light.
The two cavity mirrors introduce Dirichlet boundary conditions $\phi=0$ at the points $x=x_l$ and $x=x_r$, with $L=x_r-x_l$. Quantizing the field in Minkowski coordinates, we obtain a discrete set of bosonic cavity modes with mode functions
\be
u_n(t,x)=\frac{1}{\sqrt{\pi n}}\sin{\lr\omega_n(x-x_l)\rr}e^{-i\omega_nt}\label{minkowskifunction}
\ee
and frequencies $\omega_n=\pi n\,c/L,\hspace{3pt}n=1,2,...$.

An observer moving with constant proper acceleration $a$ is static in the Rindler coordinates $(\eta,\xi)$, defined by
\bea
x&=&\frac{c^2}{a}e^{a\xi/c^2}\cosh{\lr a\eta/c\rr},\\
t&=&\frac{c}{a}e^{a\xi/c^2}\sinh{\lr a\eta/c\rr}.
\eea
In these coordinates, the wave equation takes the same form as in equation (\ref{KGeqn}). The mirrors introduce Dirichlet boundary conditions at the points $\xi=\xi_l$ and $\xi=\xi_r$ separated by a distance $L'=\frac{c^2}{a}\arctanh{\lr \frac{aL}{2c^2}\rr}$ with respect to Rindler position $\xi$, corresponding to a proper distance $L$. Quantizing the field in Rindler coordinates gives rise to a set of bosonic cavity modes with mode functions
\be
v_m(\eta,\xi)=\frac{1}{\sqrt{\pi m}}\sin{\lr\Omega_m(\xi-\xi_l)\rr}e^{-i\Omega_m\eta}\label{rindlerfunction}
\ee
and frequencies $\Omega_n=\pi n\,c/L',\hspace{3pt}n=1,2,...$.
\newline
\\
During Rob's trip, the state in Alice's cavity will simply undergo free time-evolution in the lab frame.
To relate the inital and final states in Rob's cavity, we use Bogoliubov transformation techniques \cite{bruschi}. Before the trip, the modes in the cavity are described by a set of annilhilation and creation operators, $a_n$ and $a_n^{\dagger}$, satisfying the canonical commutation relations $[a_m,a_n^{\dagger}]=\delta_{mn}$. The modes in the cavity after the trip are similarly described by another set of operators, $b_n$ and $b_n^{\dagger}$, satisfying similar commutation relations. These two sets are related by a Bogoliubov transformation, defined by
\be
b_m=\sum_n\lr A_{mn}^*a_n-B_{mn}^*a_n^{\dagger}\rr.
\label{bogotransf}
\ee
The Bogoliubov coefficients $A_{mn}$ and $B_{mn}$ are functions of the trajectory parameters $a$, $t_a$ and $t_i$ and the proper length $L$ of the cavity. We compute the coefficients analytically as power series expansions in the dimensionless parameter $h\equiv aL/c^2$ (see Appendix \ref{sec:appendix}).
\newline
\\
The first mode in each cavity is prepared in a coherent state, with vacuum in the higher modes. Free time-evolution of a coherent state corresponds to a phase rotation. Since the proper length of the cavity is preserved throughout the trip, that is true also for the mode frequencies. Thus, we can relate the accumulated phase shift during the trip to an elapsed proper time by simply dividing with the frequency of the first mode.

The state in Alice's cavity will transform only by a phase rotation. Knowing the Bogoliubov coefficients, we can in principle fully determine the final state in Rob's cavity. We are, however, only interested in the phase shift $\theta$ of the first mode, given by (see Appendix \ref{sec:appendix})
\be
\tan{\theta}=\frac{-\text{Im}\lr A_{11}-B_{11}\rr}{\text{Re}\lr A_{11}- B_{11}\rr}.\label{phaseshift}
\ee

\section{Experimental implementation}
\begin{figure}[t]
\centering
\includegraphics[width=0.9\columnwidth]{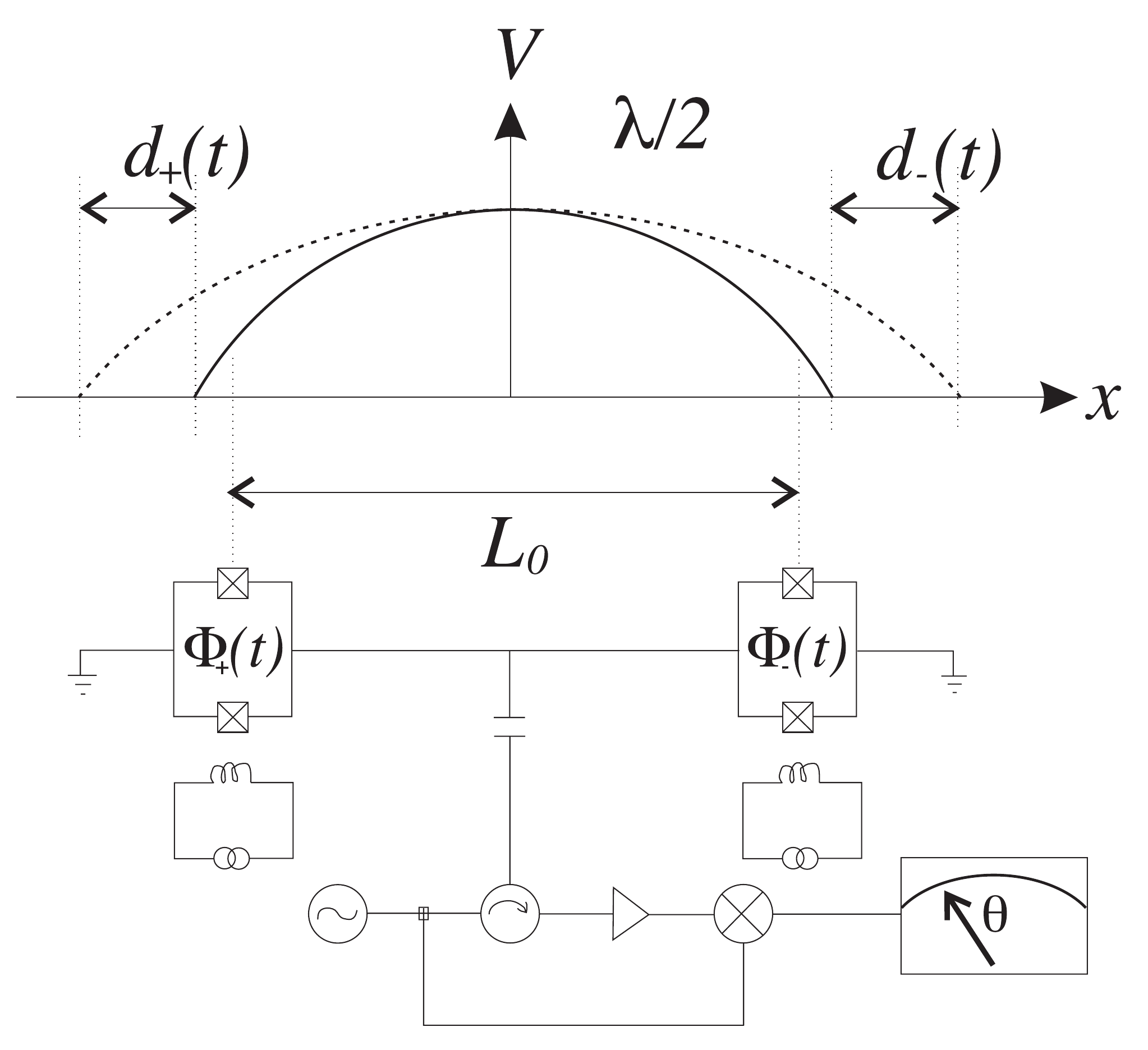}
\caption{\textbf{Experimental setup.} Above: The flux tuned SQUIDs generate time-dependent boundary conditions for the cavity field, equivalent to Dirichlet boundary conditions at different effective positions. The external fluxes $\Phi_{\pm}(t)$ correspond to effectively moving the boundary conditions the distances $d_{\pm}(t)$. Below: Sketch of the circuit setup. The signal from a coherent microwave source is used to represent Alice's clock and to fill Rob's cavity with photons. When Rob's cavity has been filled a set of travels is performed by flux tuning the two SQUIDs, using the external magnetic fluxes $\Phi_-(t)$ and $\Phi_+(t)$. After the trips, the field in the cavity leaks out and is down converted by a mixer using Alice's clock. A phase difference between the two clocks is then detected as a dc change at the output of the mixer. } 
\label{fig:setup}
\end{figure}

As already suggested in \cite{friis}, the cQED setup used to verify the dynamical Casimir effect \cite{wilson} can be expanded to simulate relativistically moving 1D cavities. A superconducting coplanar waveguide supports a $1+1$-dimensional electromagnetic field. Terminating the waveguide through a superconducting quantum interference device (SQUID) generates a Dirichlet boundary condition for the field at some effective distance from the SQUID itself. Now, by modifying the external magnetic flux through the SQUID, this effective distance can be tuned. Thus, the boundary condition becomes that of a moving mirror. Using two SQUIDs, we can construct a cavity where both mirrors can be moved along arbitrary and independent trajectories. In particular, these trajectories can be chosen so that the relativistic motion of a cavity with constant proper length is simulated (see figure \ref{fig:setup}).

To realize the twin paradox cavity trajectory described above in a cQED setup, there are several experimental constraints to take into account. The mirrors can effectively be displaced a few millimeters, while the length of the cavity itself is around a centimeter. Thus, Rob will be a ``shaking twin" rather than the usual twin going to another solar system. For such a short trip, the relative phase shift between the cavities is very small. We can, however, repeat the same trip many times in order to accumulate a larger relative phase. The limit on the number of times this can be done is set by the lifetime of the field excitations in the cavity. The cavity can then be filled again with photons so that the measurement can be repeated arbitrarily many times. Moreover, the plasma frequency of each SQUID must be larger than all the other frequencies involved, limiting the effective velocities and accelerations.

As an example of what can be achieved in the cQED setup, see figure \ref{fig:phaseplot}. In this example we let the microwave source play the role of Alice's cavity. Assuming state of the art arbitrary waveform generators to source the fluxes through the SQUID loops, it should be possible to make $t_a$ as small as 1 ns while still maintaining the required waveform. In this case, the effective acceleration is limited to $1.7\times10^{15}\, m/s^2$ if the maximal allowed flux through the SQUIDs is not to be exceeded. For a standard cavity length of $1.1$ cm, this corresponds to $h=1.3\times 10^{-3}$. For the parameter values listed above, and with $t_i=0$, we predict relative phase shifts of up to 130 degrees, which is detectable. This scenario would correspond to an effective cavity displacement of 1.7 mm and a maximal velocity of $1.4\%$ of the speed of light. With $t_t=4$ ns and the trajectory being repeated 500 times, the total travel time is $2\mu$s. The time difference related to the relative phase shift agrees with what we would obtain if Alice and Rob were instead carrying pointlike ideal clocks. Thus, we can conclude that it is challenging but possible to simulate the twin paradox scenario in a cQED setup.
 
\section{Comparison to pointlike clock}
Our cavity clock agrees very well with a pointlike ideal clock in the parameter regime considered above. The reason for this is that we can choose a small $h$-value and still accumulate a phase shift large enough to observe. To second order in $h$, however, we start to see a discrepancy between the cavity clock and a pointlike one. This difference is due to both the finite extension of the cavity and the fact that non-uniform acceleration leads to mode-mixing and particle creation, eventually resulting in a different phase shift for the first mode. First, neglecting the latter effects, we note that a cavity clock differs from a pointlike one during acceleration only. During an accelerated segment, the proper time elapsed according to the cavity clock is
\be
\tau^a_{\text{cav}}=\frac{\theta_a}{\omega_1}=\frac{L}{c}\frac{\arcsinh{\lr at_a/c\rr}}{2\arctanh{\lr h/2\rr}}\label{taucav},
\ee
while the corresponding expression for the pointlike clock, obtained by integrating Rob's proper time over the trajectory, is
\be
\tau^a_{\text{point}}=\frac{c}{a}\arcsinh{\lr at_a/c\rr}\label{taupoint}.
\ee
Thus, the ratio of the proper times is given by
\be
\frac{\tau^a_{cav}}{\tau^a_{point}}=\frac{(h/2)}{\arctanh{\lr h/2\rr}}=1-\frac{h^2}{12}+\mathcal{O}(h^4),\label{comparison}
\ee
which is smaller than one and decreases with $h$. This means that an extended clock is slowed down during acceleration. The larger the clock, the slower its rate. The effects of mode-mixing and particle creation depend only on changes in acceleration and are encoded in the second-order terms of the Bogoliubov coefficients (\ref{A11}) and (\ref{B11}).

\begin{figure}[t]
\centering
\includegraphics[width=0.9\columnwidth]{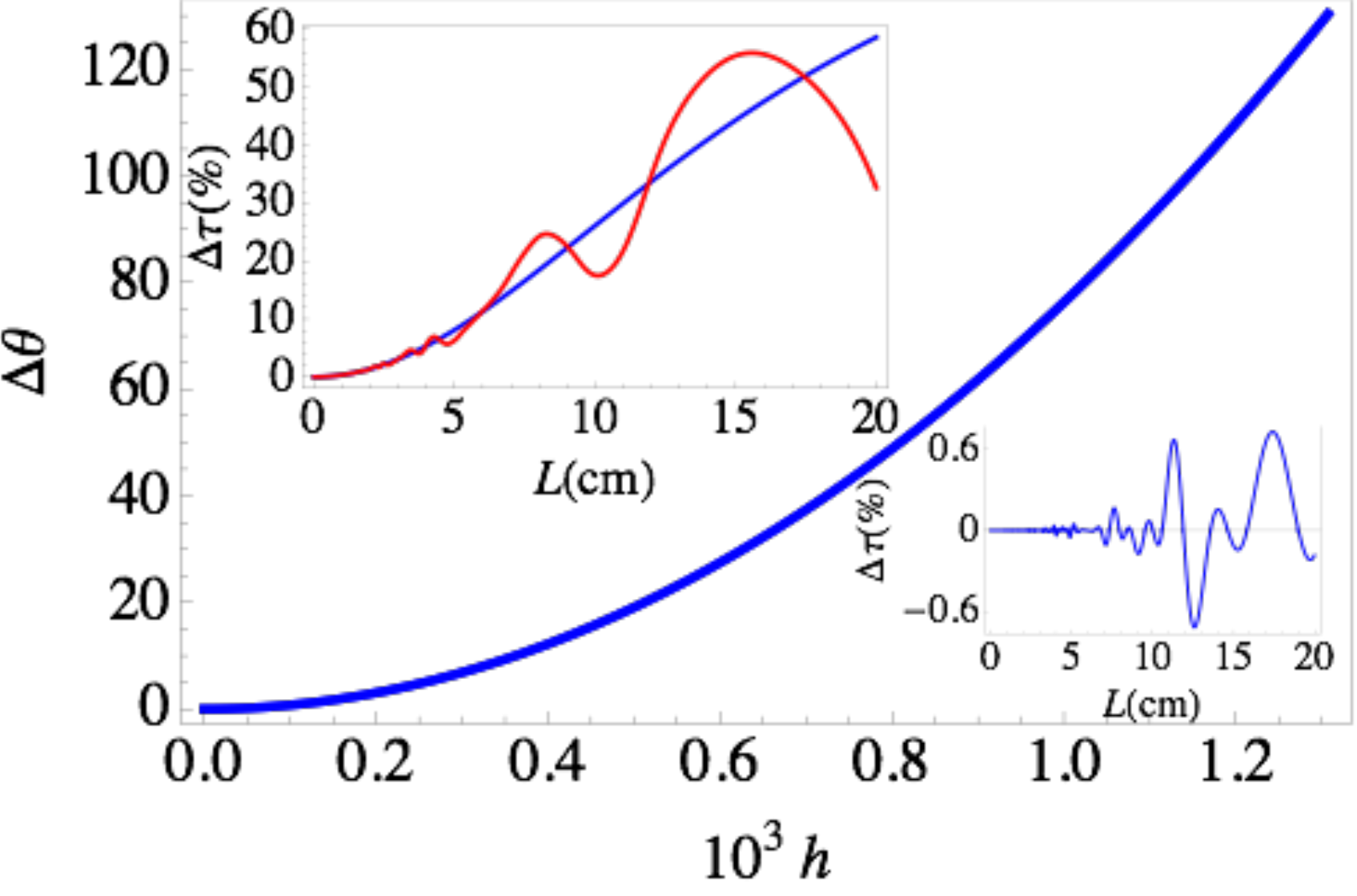}
\caption{\textbf{Time dilation.} Relative phase shift between Rob's and Alice's cavities in an experimentally feasible regime. The parameter values used are $t_a=1$ ns, $t_i=0$ and $L=1.1$ cm, leading to an effective cavity displacement of $1.7$ mm and a maximal velocity of $0.014 c$. With $t_t=4$ ns and the trajectory being repeated 500 times, the total travel time is $2\mu$s. Left inset: Difference between the time dilation shown by the cavity clock and a pointlike clock as a function of $L$, normalized to the total time dilation between Alice and Rob. The blue (red) curve is excluding (including) the effects of mode-mixing and particle creation. Right inset: Difference in time dilation between the cases with and without particle creation, again normalized to the total time dilation. The parameter values used in the inset plots are $t_a=1$ ns, $t_i=0$ and $a=1.7\times 10^{15}\, m/s^2.$}
\label{fig:phaseplot}
\end{figure}

In order to observe the higher order effects, we need to use larger $h$-values. In earlier cQED experiments \cite{wilson}, effective accelerations up to $4\times10^{17}\, m/s^2$ have been achieved. With such accelerations, though, the time $t_a$ would have to be very short, making the effective motion of the mirrors difficult to control. What we can do instead is to increase $h$ by using larger cavities. The inset plots of figure \ref{fig:phaseplot} show the shift in time dilation due to the different effects, as a function of $L$. In order to observe the effects of clock size, we can choose $L=6$ cm, which is easily realizible in the cQED setup. In this regime, the clock size is clearly the dominant effect and would contribute with an additional phase shift of 3 degrees, which is possible to resolve in the measurement stage. For even larger $L$, the other effects start to become important, with mode-mixing being the dominant one. This can be clearly seen in the right inset, where we plot the difference in phase shift between the cases with and without particle creation.

\section{Conclusions}
In conclusion, we have shown that using state-of-the-art superconducting circuit technology, the twin paradox can be demonstrated in a ground-based experiment at velocities approaching 1.4\% of the speed of light. Using the phase of a coherent state inside the cavity as a clock pointer, we find that the time dilation produces a phase shift of up to 130 degrees, which is clearly in the measurable regime.  We also note that at high accelerations the extension of the clock becomes relevant: time dilation increases with the clock's spatial dimension. This opens up an avenue for the experimental exploration of the differences between a pointlike \cite{rossi,bailey,chou} and a physically extended clock.  In the near future, we foresee that other quantum effects on clock accuracy and time dilation can be explored using squeezed cavity states.
By analysing for the first time the twin paradox in a framework of quantum field theory with boundary conditions corresponding to relativistic motion, we are able to study theoretically the interplay of quantum effects, such as the dynamical Casimir effect, in a paradigmatic relativistic effect such as time dilation. In this way we take a step further in our knowledge on the overlap between quantum theory and relativity.
\appendix
\section{Bogoliubov coefficients}
\label{sec:appendix}
To determine the Bogoliubov coefficients $A_{mn}$ and $B_{mn}$ in equation (\ref{bogotransf}) we use techniques developed in \cite{bruschi}. The Bogoliubov coefficients relating the modes of an inertial observer to those of a uniformly accelerating observer are, expressed as Klein-Gordon inner products \cite{birrelldavies},  $\alpha_{mn}=(v_m,u_n)$ and $\beta_{mn}=-(v_m,u_n^*)$, where $u_n$ and $v_m$ are given by (\ref{minkowskifunction}) and (\ref{rindlerfunction}). $\alpha_{mn}$ and $\beta_{mn}$ account for mode-mixing and particle creation, respectively. The resulting integrals cannot be evaluated in terms of elementary functions, but to second order in $h\equiv a L/c^2$ we can write the coefficients as
\bea
\alpha_{mn}&=&\alpha_{mn}^{(0)}+\alpha_{mn}^{(1)}h+\alpha_{mn}^{(2)}h^2,\label{bogoalpha}\\
\beta_{mn}&=&\beta_{mn}^{(0)}+\beta_{mn}^{(1)}h+\beta_{mn}^{(2)}h^2,\label{bogobeta}
\eea
with
\bea
\alpha_{nn}^{(0)}&=&1,\hspace{3pt}\alpha_{nn}^{(1)}=0,\hspace{3pt}\alpha_{nn}^{(2)}=-\frac{\pi^2n^2}{240},\nonumber\\
\alpha_{mn}^{(0)}&=&0,\hspace{3pt}\alpha_{mn}^{(1)}=\sqrt{mn}\frac{(-1)^{m-n}-1}{\pi^2\lr m-n\rr^3},\hspace{5pt}    m\neq n\nonumber\\
\alpha_{mn}^{(2)}&=&\sqrt{mn}\frac{\lr(-1)^{m-n}+1\rr\lr m+2n\rr}{2\pi^2\lr m-n\rr^4},\hspace{5pt}m\neq n\nonumber\\
\beta_{mn}^{(0)}&=&0,\hspace{3pt}\beta_{mn}^{(1)}=\sqrt{mn}\frac{1-(-1)^{m-n}}{\pi^2\lr m+n\rr^3},\hspace{5pt}m\neq n\nonumber\\
\beta_{mn}^{(2)}&=&\sqrt{mn}\frac{\lr-(-1)^{m-n}-1\rr\lr m-2n\rr}{2\pi^2\lr m+n\rr^4}\label{bogb2}.
\eea
During each accelerated segment of the trip, the fundamental mode of the cavity aquires the phase
\be
\theta_a=\frac{\pi\arcsinh{\lr at_a/c\rr}}{2\arctanh{\lr h/2\rr}},
\ee
while for an inertial segment the corresponding phase shift is
\be
\theta_i=\pi ct_i/(\gamma L),
\ee
$\gamma=\sqrt{(at_a/c)^2+1}$ being the Lorentz factor during the inertial motion. By composing transformations described by equations (\ref{bogoalpha})-(\ref{bogb2}) and their inverses, with appropriate Rindler and Minkowski time-evolution phase transformations in between, we can find the Bogoliubov coefficients relating the cavity modes before and after the trip. Only terms up to second order in $h$ are kept.

Acting with the Bogoliubov transformation on the vector of first moments of the cavity state and tracing out the higher modes, the expression in equation (\ref{phaseshift}) is obtained for the phase shift of the first mode, provided that the initial phase is zero. The explicit expressions for the relevant coefficients are
\bea
A_{11}&=&\lr 1+6\alpha_{11}^{(2)}h^2\rr e^{i(4\theta_a+2\theta_i)}\nonumber\\
&&+h^2\sum_{k=2}^{\infty}\lr\alpha_{k1}^{(1)}\rr^2\times\nonumber\\
&&\ls 2e^{(k+3)i\theta_a+2i\theta_i}+2e^{2(k+1)i\theta_a+(k+1)i\theta_i}\right.\nonumber\\
&&\left.-2e^{(3k+1)i\theta_a+(k+1)i\theta_i}-2e^{(3k+1)i\theta_a+2ki\theta_i}\right.\nonumber\\
&&\left.+2e^{(k+3)i\theta_a+(k+1)i\theta_i}-2e^{4i\theta_a+(k+1)i\theta_i}\right.\nonumber\\
&&\left.+e^{2(k+1)i\theta_a+2i\theta_i}+e^{4ki\theta_a+2ki\theta_i}+e^{2(k+1)i\theta_a+2ki\theta_i}\rs\nonumber\\
&&-h^2\sum_{k=2}^{\infty}\lr \beta_{k1}^{(1)}\rr^2\times\nonumber\\
&&\ls 2e^{(-k+3)i\theta_a+2i\theta_i}+2e^{2(-k+1)i\theta_a+(-k+1)i\theta_i}\right.\nonumber\\
&&\left.-2e^{(-3k+1)i\theta_a+(-k+1)i\theta_i}
-2e^{(-3k+1)i\theta_a-2ki\theta_i}\right.\nonumber\\
&&\left.+2e^{(-k+3)i\theta_a+(-k+1)i\theta_i}-2e^{4i\theta_a+(-k+1)i\theta_i}\right.\nonumber\\
&&\left.+e^{2(-k+1)i\theta_a+2i\theta_i}+e^{-4ki\theta_a-2ki\theta_i}+e^{2(-k+1)i\theta_a-2ki\theta_i}\rs,\nonumber\\
\label{A11}
\eea
\bea
B_{11}&=&2ih^2\beta_{11}^{(2)}\ls \sin{\lr 4\theta_a+2\theta_i\rr}-\sin{\lr 2\theta_a+2\theta_i\rr}+\sin{\lr 2\theta_a\rr}\rs\nonumber\\
&&+2ih^2\sum_{k=2}^{\infty}\lr\alpha_{k1}^{(1)}\beta_{k1}^{(1)}\rr\times\nonumber\\
&&\ls \sin{\lr \lr4\theta_a+2\theta_i\rr k\rr}-2\sin{\lr \lr3\theta_a+2\theta_i\rr k\rr}\cos{\lr\theta_a\rr}\right.\nonumber\\
&&\left.-2\sin{\lr \lr3\theta_a+\theta_i\rr k\rr}\cos{\lr\theta_a+\theta_i\rr}+\sin{\lr \lr2\theta_a+2\theta_i\rr k\rr}\right.\nonumber\\
&&\left.+2\sin{\lr \lr2\theta_a+\theta_i\rr k\rr}\cos{\lr\theta_i\rr}+\sin{\lr 2\theta_ak\rr }\right.\nonumber\\
&&\left.+2\sin{\lr \lr\theta_a+\theta_i\rr k\rr}\cos{\lr 3\theta_a+\theta_i\rr}\right.\nonumber\\
&&\left.+2\sin{\lr \theta_ak\rr}\cos{\lr 3\theta_a+2\theta_i\rr}
-2\sin{\lr\theta_ik\rr}\cos{\lr2\theta_a+\theta_i\rr}\rs.\nonumber\\
\label{B11}
\eea

\section*{Acknowledgments}
We thank Barry Sanders and Vitaly Shumeiko for valuable discussions.  J. L, G. J, I.-M. S. and P. D. would like to acknowledge funding from the Swedish Research Council and from the EU through the ERC. I. F. and C.S acknowledges support from EPSRC (CAF Grant No. EP/G00496X/2 to I. F.).  A. D. is funded by National Science
Center, Sonata BIS grant 2012/07/E/ST2/01402.

\end{document}